\documentclass[preprint,prb,showpacs,preprintnumbers,amsmath,amssymb]{revtex4}

\usepackage{graphicx}
\usepackage{dcolumn}
\usepackage{bm}
\def\MSB{$\mu$\textsc{squid}~}
\def\MSA{$\mu$\textsc{squid}}

\begin{document}

\preprint{APS/123-QED}

\title{Direct observation of vortices in an array of holes at low 
temperature: temperature dependance and first visualization of localized superconductivity}

\author{C. Veauvy}
\author{K. Hasselbach}
\affiliation{CRTBT-CNRS, associ\'{e} \`{a} l'Universit\'{e} Joseph Fourier, \\ 25 avenue des Martyrs, BP 166 X,38042 Grenoble, France.}%

\author{D. Mailly}%
\affiliation{LPN-CNRS, \\Route de Nozay, 91460 Marcoussis, France.}%

\date{\today}

\begin{abstract}
A scanning micro superconducting quantum interference device (\MSA) microscope is used to directly image vortices in a superconducting Al thin film. We observe the temperature dependence of the vortex distribution in a regular defect (hole) array patterned into the Al film. The first direct observation of the localized superconducting state around the holes is shown as well as the effect of the hole size on nucleation of the superconducting state.
\end{abstract}

\pacs{74.25.Qt, 74.25.Op, 74.78.Db, 85.25.Dq}

\maketitle

\section{Introduction}

The mixed state of a type II superconductor is typified by the presence of vortices, flux tubes containing one quantum of flux ($\Phi =h/2e$). 
The vortex core of radius  $\xi$  contains only normal electrons and the magnetic field of the vortex diminishes over a length  $\lambda$ perpendicular to the vortex axis.
The static and dynamic properties of the vortices are dependent on the structure of the superconductor, the defects present in the material, the temperature and the applied magnetic field to name the most important. Besides its academic interest, an understanding of the magnetic properties of the superconducting mixed state is a major issue for technological application of the superconductors since the motion of the vortices in the presence of an electrical current induces a voltage drop and the appearance of electrical resistance. Therefore, there is considerable interest in developing the ability to control the pinning of vortices at a fixed position inside the superconductor.

The efficiency of a pinning center depends in part on the coherence length ($\xi$) and the penetration depth ($\lambda$) of the superconductor. Recent improvements in micro fabrication techniques have made possible the realization of superconductor films with pinning center arrays where the size of each defect and the distance between them are comparable to $\xi$ and $\lambda$. Many different sorts of regular defect arrays have been fabricated using, for example, periodic modulation of film thickness \cite{daldini}, patterned holes (or antidots) \cite{fiory,baert,bezryadin} and arrays of magnetic dots \cite{vanbael,hoffmann}. In the case of the hole arrays, bulk measurements \cite{fiory,baert2,moshchalkov} and 2D magnetic imaging \cite{bending,bezryadin,bending2} have been performed showing a commensurability effect between the vortex array and the defect array as well as exploring the role of the defect size and the vortex saturation number for a given hole. 
For example Lorentz microscopy \cite{tonomura} was used to dynamically image the penetration of vortices inside a superconducting thin film with a regular hole array. Most of these experiments have probed the dynamics of the vortices by changing the applied magnetic field.

In the present experiment, vortices were directly observed in an Al superconducting thin film with a regular hole array. The  presence of the holes allows the observation of two kinds of vortices. The first kind appears as vortices strongly pinned in holes: the flux in the holes is quantized due to the fluxoid quantization. The second kind of vortices is located at the interstices between the holes: they are Abrikosov vortices and will be called interstitial vortices. These vortices are free to move whenever they overcome the weak pinning at the impurities of the Al layer.

In the first part of this paper, the experimental setup and the sample will be briefly presented. Then, we will show how vortices arrange in the presence of the hole array for different temperatures. Finally, the first direct observation of the localized state at the hole surface will be shown, and we will explore the effect of the hole size on the nucleation of superconductivity.

\section{The experimental setup}

The present experiment was carried out with a \MSB force microscope (\MSA -FM). This microscope is  based on a new technique associating a magnetic sensor, the \MSA, and a force sensor. The magnetic flux is detected by a \MSB patterned by electron beam lithography and consisting of a 1.2$\mu$m diameter loop interrupted by two Josephson junctions. The \MSB is implemented on a quartz tuning fork which is used as a mechanical resonator to detect surface forces and is part of a feedback loop to maintain the distance between \MSB and sample surface constant. The \MSB is scanned at about one micron from the sample surface to provide 2D magnetic imaging. The integration of the microscope into a dilution refrigerator allows us to image at temperatures below 1 K and therefore offers a largely unexplored temperature range in magnetic imaging. The sample is thermalized independently of the \MSB so we can study the sample at different temperatures without disturbing the \MSA. A Helmholtz coil located at the outside of the dilution refrigerator insures a homogeneous magnetic field on the whole sample. A more detailed description of this microscope has been published recently in the literature \cite{klaus_cecile}. During imaging the microscope has a  magnetic spatial resolution better than 2 micrometers and a
magnetic flux sensitivity of $10^{-3} \frac{\phi_0}{\sqrt{Hz}}$, where $\phi_0 =   \frac{h}{2e} $ is the flux quantum.

\section{The sample}

\begin{figure}[h!tb]
\includegraphics[width=8cm]{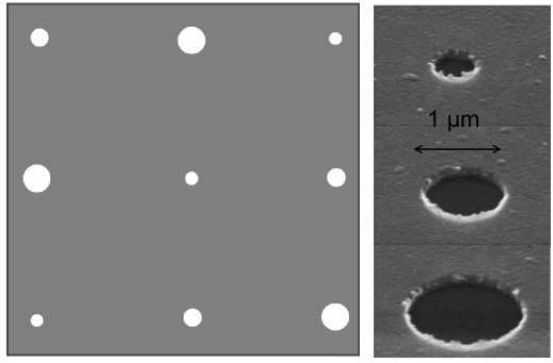}
\caption{Schematic view of the hole array unit cell and SEM images of the holes made in an Al thin film. The holes are $10\mu$m apart and they have three different diameters: $0.5\mu$m, $1\mu$m and $1.5\mu$m. The entire hole array size is $3 \times 3 mm^{2}$.}
\label{sample}
\end{figure}

In the present experiment, the aluminum film contained an array of holes produced by lift-off electron-beam lithography with positive UV3 photoresist. The holes formed a two dimensional square array with spacing $d=10\mu$m and within the array the holes have 3 different diameters: $0.5\mu$m, $1\mu$m and $1.5\mu$m, as shown in Fig. \ref{sample}. Sizes in the neighborhood of $1\mu$m were chosen because this optimizes the flux coupling between the flux at the hole and the \MSA.

All the measurements were carried out on an aluminum film of thickness 170 nm prepared by thermal evaporation of pure aluminum in a vacuum of $10^{-7}$ mbar. AFM allowed the surface state characterization of the film and indicated roughness of the order of some tens of nanometer. Even though bulk aluminum is a type I superconductor, in thin film form it behaves like a type II superconductor. The limit between a type I and a type II superconductor is given by the Ginzburg-Landau parameter \cite{degennes}, $\kappa=\frac{\lambda}{\xi}$: for $\kappa<\frac{1}{\sqrt{2}}$, the superconductor is type I, otherwise it is type II. In a thin film, the effective penetration depth, $\lambda_{eff}$, is a function of the film thickness \cite{degennes} and becomes $\lambda_{eff}=\frac{\lambda^{2}}{d}$. Therefore the parameter $\kappa$ becomes $\kappa=\frac{\lambda^{2}}{\xi d}$ and below a critical film thickness, a type I superconductor in bulk form exhibits a type II behavior. So for temperatures below its critical temperature ($T_c$) and under magnetic field, vortices appear in such a thin film. Interactions between vortices are mainly governed by the coherence length $\xi$ of the material. In the case of aluminum, $\xi$ is rather large (few hundred of nanometers). Therefore, as we will see, even with a small external magnetic field applied, one can observe the transition from a single object state where vortices behave like independent objects (low temperature where $\xi$ is smaller) to a collective state where vortices interact with each other ($T \rightarrow T_{c}$ where $\xi \rightarrow + \infty$).

An estimate of the sample's coherence length can be obtained by measuring the temperature dependence of the critical field  $H_{c2}$(T) of the superconducting film close to $T_{c}$.

\begin{figure}[h!tb] 
\includegraphics[width=7cm]{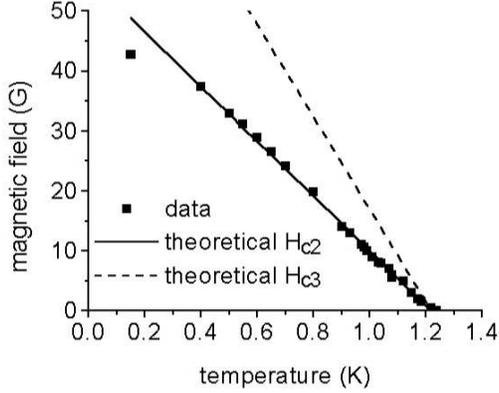}
\caption{Temperature dependance of the critical magnetic field $H_{c2}$ (dots) determined by transport measurement. The full line represents $H_{c2}(T)$ predicted by Ginzburg-Landau theory using a coherence length $\xi(0)=240nm$. The theoretical critical field $H_{c3}$ ($H_{c3}=1.69H_{c2}$) is shown with the dashed line.}
\label{Hc2}
\end{figure}
 
The critical field vs. temperature phase diagram obtained using a standard 4 point resistance measurement is presented in Fig. \ref{Hc2}. From this H-T diagram, one can deduce $T_{c}(H=0)=1.23K$ and $H_{c2}(T=0)=43G$. Near $T_c$, there is a very good agreement between the data and the theoretical temperature dependence of $H_{c2}$ ($H_{c2}=\frac{\phi_{0}}{2\pi\xi(0)^{2}}\left(1-\frac{T}{T_{c}}\right)$) allowing a coherence length at zero temperature $\xi(0)=240nm$ to be determined. Figure \ref{Hc2} also shows the temperature dependence of $H_{c3}$ based on $H_{c3}=1.69H_{c2}$\cite{saintjames}.  The region between  $H_{c3}$ and $H_{c2}$ is the region where  the bulk sample is already in the normal state and only a superconducting sheath persists at surfaces of the sample parallel to the applied field.

\section{Temperature dependance of the vortex distribution}

\begin{figure} [h!tb]
\includegraphics[width=13.5cm]{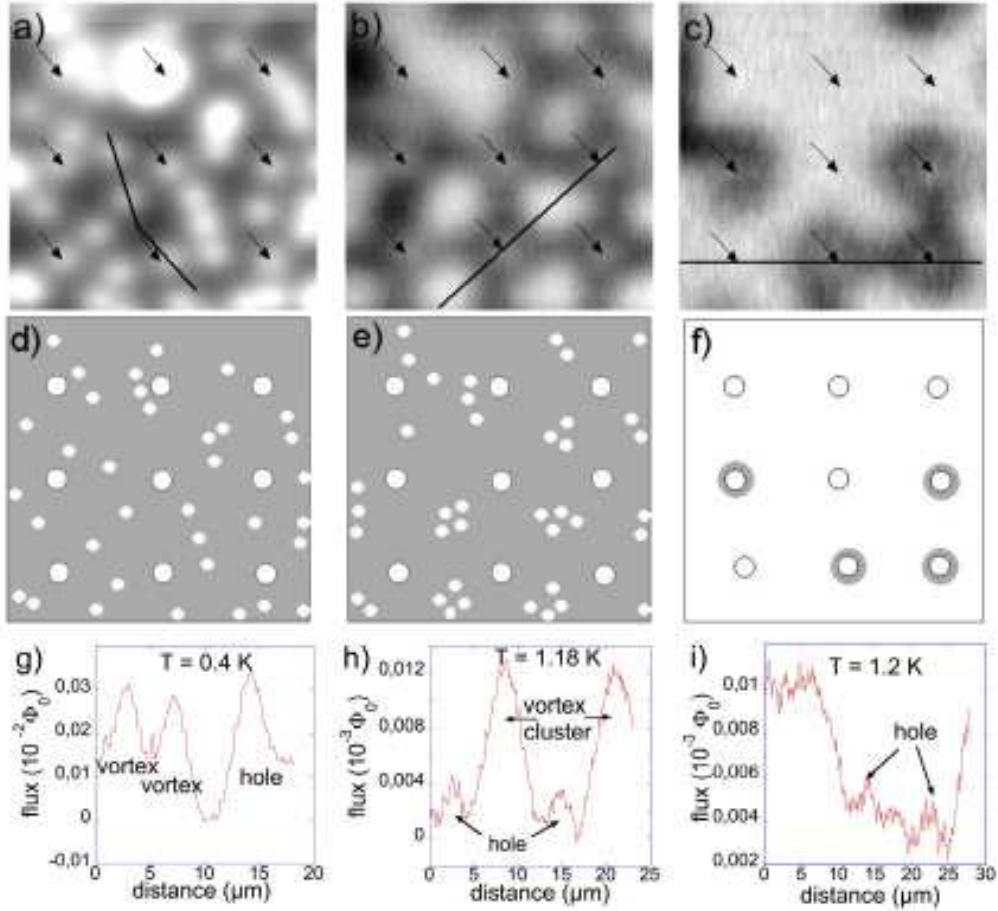}  

\caption{Images a,b,c  ($28 \times 28 \mu m^{2}$) realized by \MSA-FM after cooling under a magnetic field of $1.14G$ down to three different temperatures: (a) $T=0.4K$ 
(b) $T=1.18K$ and (c) $T=1.2K$. The arrows point out the location of the hole array in the Al film. Figs. d,e,f) present, schematically, the flux and vortex configuration (white) in registry with the underlying hole array.  The open circles represent the holes and the dots the flux, superconducting regions are gray. Figs. g,h,i) show line profiles along the dark lines of Figs. a,b,c) in  units of $ {\phi_0}$ of the \MSA.}
\label{temp_dependance}
\end{figure}

The experiment consists in the imaging of the magnetic flux passing through the \MSB loop as it is scanned over the sample surface.  
Measurements presented here are done as a function of temperature and in a magnetic field of 1.14G. Figure \ref{temp_dependance}(a,b,c) shows the magnetic mapping of a $28 \times 28\mu m^{2}$ sample area for 3 different temperatures (0.4K, 1.18K and 1.2K respectively). On the images, white areas correspond to high magnetic flux and reveal the presence of vortices. The measured vortex density (Fig. \ref{temp_dependance}a) agrees with the expected vortex density (45 vortices in $28 \times 28 \mu m^{2}$). We have indicated the location of the holes in the Al film with arrows, the Figs. \ref{temp_dependance} d,e,f) schematize the images a,b,c). 

\begin{figure} [h!tb]
\includegraphics[width=4.5cm]{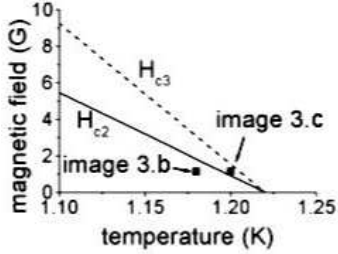}
\caption{H-T diagram showing $H_{c2}$ and $H_{c3}=1.69H_{c2}$ temperature dependence and the temperature and magnetic field where images \ref{temp_dependance}b) and \ref{temp_dependance}c) were acquired.}
\label{HT_diagram_detail}
\end{figure}

When the sample is cooled down in an applied field of $1.14G$ to low temperature ($T=0.4K<<T_{c2}$) (cf Fig. \ref{temp_dependance}a) one observes a flux quantum in each hole and the configuration of the interstitial vortices is random. At such temperature, the characteristic lengths, $\lambda$ and $\xi$, are relatively small (typically 200-300 nm) compared to the size of the hole array, thus  vortices are not sensitive to the periodic aspect of the hole array but are mainly affected by the nearest hole. This state is called single object state. The repulsive force between Abrikosov vortices, being inversely proportional to $\lambda$ is maximum at low temperature.
Moreover, the small size of the vortices and the low thermal energy allow them to be easily pinned by impurities on the aluminum layer which are of small size (AFM imaging of the sample shows a corrugation of some tens of nanometers). Therefore vortices are spread over the aluminum layer following the random location of the impurities of the superconductor. This state is quite stable because two consecutive images (corresponding to a 20 minute time interval) show the same vortex configuration.

Fig. \ref{temp_dependance}b) is an image at $T=1.18K$. At this temperature, the sample is still in the mixed state: the aluminum layer is superconducting and pierced by vortices. The interstitial Abrikosov  vortices are ordering between the flux anchored at the holes. The quantized flux in the holes generates currents around the hole edge, establishing a repulsive interaction with the interstitial vortices. The interstitial vortices are accumulating at the central site of the square hole lattice. As interstitial vortices are seen to arrange freely we deduce that the activation energy close to $T_{c2}$ is sufficient to unpin the vortices from the pinning sites at the film surface. At this temperature, the vortex characteristic lengths $\lambda$ and $\xi$ which are proportional to $\sqrt{\frac{1}{1-\frac{T}{T_{c2}}}}$, are large compared to the nm scale rugosity of the Al film and their length is a large fraction of the period of the hole array.  Thus the vortices are sensitive to the presence of the artificial hole array and tend to localize in the interstices of this array to minimize the energy of the system. Besides,
the repulsive interaction between vortices is small near $T_{c2}$ and the vortices can be closer to each other. Therefore vortices are in a collective state where a long range organization of interstitial vortices appears mediated by the flux in the hole array. Similar ordering has been observed for other flux densities, e.g; single Abrikosov vortices sit in the center of a cell of the hole array.
Flux counting, using Fig. \ref{temp_dependance}e), suggests that the interstitial sites carry three to four flux quanta, this is also supported by the line profile of Fig. \ref{temp_dependance}h). 
With the present spatial magnetic and temporal resolution of our microscope, we cannot conclude if interstitial vortices still carry a single quantum of flux ($\phi_{0}$) or if they carry multiquanta. Nevertheless the positioning of the vortices at the interstices of the array imaged with the microscope is in very good agreement with the minimum of potential calculated by I.B. Khalfin and al \cite{khalfin}, in the case of a periodical lattice of columnar defects in a type II superconductor.

When the temperature is increased to $T=1.2K$ (Fig. \ref{temp_dependance}c), one observes the general disappearance of magnetic contrast. Magnetic contrast prevails only around some holes (we will discuss below the fact that all the holes do not show the same magnetic contrast). This indicates that the superconductivity nucleates at the hole surface while the aluminum layer transits to the normal state:
only a thin loop (size of the order of $\xi$) around the holes is still superconducting. These observations are consistent with the experimental conditions of temperature ($1.2K$) and magnetic field ($1.14G$) in the H-T diagram (Fig. \ref{HT_diagram_detail}): we conclude that the sample is in the zone of localized superconductivity.
For temperature higher than 1.23K, the sample appears magnetically uniform (not shown): it is in the normal state and the magnetic field penetrates the aluminum layer uniformly.

\section{Observation of localized superconductivity}

Superconductivity is localized in a plane  of thickness $\xi$ at the surface of a	superconductor, if the applied magnetic field is parallel to this plane and with an amplitude between 1.69$H_{c2}$
=  $H_{c3}$ and $H_{c2}$.  As the  magnetic field penetrates the superconductor in the bulk, the surface region carries in its section a circulating current over a thickness of the order of $\xi$. These surface currents cannot screen out the field inside the specimen \cite{Abrikosov}. 
The current density is high at the edge, reverses sign and levels off to zero after less then 3$\xi$ inside the specimen. The integral over the current density is zero in the direction perpendicular to the surface   \cite{Fink}. 
When the specimen is no longer singly connected but becomes multiply connected, (by connecting the ends of the superconductor e.g.) the current distribution must rearrange in order to establish a single valued phase around the hole.  For fields between $H_{c2}$
and  $H_{c3}$  this state may carry a magnetic moment \cite{Fink1980} and superconductivity is localized.

A similar situation can be found in a planar film with holes. Localized superconductivity is expected to arise when the field is applied  perpendicular to the plane of the holes i.e. parallel to the edge of the hole, with an amplitude between $H_{c2}$
and  $H_{c3}$. A superconducting annulus will appear at the edges of the holes.

\begin{figure} [h!tb]
\includegraphics[width=5cm]{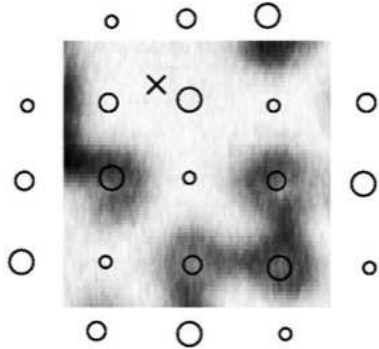}
\caption{Comparison in between the magnetic image of the Al film cooled at $1.2K$ under a magnetic field of $1.14G$ and a schematic view of the underlying hole array. The cross shows the place of a large defect in the Al film that disturbs the magnetic order. Aligned along a diagonal, the smallest diameter holes  do not expell magnetic flux anymore.}
\label{supra_surface}
\end{figure}

If we consider the image presenting the localized state (Fig. \ref{temp_dependance}a), as we have already noted, the holes do not seem to be equivalent since superconductivity is nucleated only at the surface of some holes. This observation shows the role of hole size on the nucleation of the superconductivity. Figure \ref{supra_surface} shows a schematic view of the hole configuration in registry with the distribution of magnetic flux. Magnetic contrast is observed only at the surface of the largest holes whereas the smallest holes are totally invisible: superconductivity is only present around the larger hole sizes. 
For localized superconductivity, flux quantization is not satisfied as the shielding is incomplete. This leads to less magnetic contrast at the holes Fig. \ref{temp_dependance}a,g).   In Fig. \ref{supra_surface}, the cross shows the presence of a defect in the aluminum layer disturbing the localized superconductivity at two site.  This defect is also visible on the three images of Fig. \ref{temp_dependance}). We conclude from Fig. \ref{supra_surface} that the critical magnetic field is a function of the geometric size of the hole. This hole size dependence of $H_{c3}$ has been theoretically \cite{buzdin} and numerically\cite{meyers} shown. This work demonstrated how $H_{c3}$ tends to $H_{c2}$ when the size of defect decreases and tends to $1.69H_{c2}$ when this size diverges.

\section{Conclusion}
In conclusion, we have presented the first direct observation of vortices in Al. We have been able to study the temperature dependance of the vortex distribution in a regular defect array showing the vortex transition from a collective state (T near $T_{c}$) to a single object state ($T<<T_{c}$). Finally, under special conditions of temperature and magnetic field, the first direct observation of localized superconductivity around holes and its hole size dependence were presented.

\begin{acknowledgments}
We acknowledge the support of CNRS ULTIMATEC, DGA Contract No 96 136 and CNRS NOI. The fruitful discussions with B. Pannetier, S. Buzdin, C. Meyers and R.T. Collins were greatly appreciated. The authors also thank the technical assistance of J.L. Bret and M. Grollier and the advices of H. Courtois and A. Benoit.
\end{acknowledgments}

\bibliography{apssamp}

\end{document}